\documentclass[naps,prl,preprint,notitlepage,superscriptaddress,floatfix]{revtex4-2}
\usepackage{graphicx}
\usepackage{dcolumn}
\usepackage{bm}
\usepackage{amsmath}
\usepackage{amssymb}
\usepackage{hyperref}
\begin{document}
\title{Universal Dynamic Scaling for Sub-Yield Aging and Over-Yield Rejuvenation in A Cyclically Driven Metallic Glass}
\author{Nannan Ren}
\affiliation{School of Materials Science and Engineering, Anhui University of Technology, Maanshan 243032, China}
\author{Hongfei Zhang} 
\affiliation{School of Materials Science and Engineering, Anhui University of Technology, Maanshan 243032, China}
\author{Rui Wan}
\affiliation{School of Materials Science and Engineering, Anhui University of Technology, Maanshan 243032, China}
\author{Pengfei Guan}
\email[Corresponding author: ]{pguan@nimte.ac.cn}
\affiliation{Advanced Interdisciplinary Science Research Center (AiRCenter), Ningbo Institute of Materials Technology and Engineering, Chinese Academy of Sciences, Ningbo 315201, China}
\author{Limin Wang}
\email[Corresponding author: ]{Liminwang@hebut.edu.cn}
\affiliation{State Key Laboratory of High Performance Roll Materials and Composite Forming, School of Material Science and Engineering, Hebei University of Technology, Tianjin 300401, China}
\author{Baoshuang Shang}
\email[Corresponding author: ]{shangbaoshuang@hebut.edu.cn}
\affiliation{State Key Laboratory of High Performance Roll Materials and Composite Forming, School of Material Science and Engineering, Hebei University of Technology, Tianjin 300401, China}
\date{\today}
\begin{abstract}
Whether sub-yield aging and over-yield rejuvenation in cyclically deformed amorphous solids share a common physical origin or arise from distinct microscopic mechanisms remains a fundamental yet unresolved debate. Using molecular dynamics simulations of a metallic glass, we resolve this duality by uncovering a unified dynamic clock that dictates energy landscape navigation across both regimes. We show that the transition between regimes reflects a fundamental shift in energy injection pathways:
(i) In the over-yield regime, rejuvenation is driven as a first-order response to macroscopic plastic shear work derived from Herschel-Bulkley non-linear rheological scaling.
(ii) In contrast, this macroscopic work scaling fails in the sub-yield regime, where early-stage energy landscape elevation and subsequent cyclic aging are instead governed by second-order localized elastic stress fluctuations acting as a mechanical catalyst and noise source, respectively. Intense high-rate agitation increases the escape probability from deeper potential energy minima, thereby suppressing the retention efficiency of structural relaxation. 
By identifying these localized fluctuations as the primary origin of rate-attenuated sub-yield relaxation, we establish a exact scaling identity relating the relaxation attenuation exponent to the stress fluctuation exponent. 
This work directly bridges non-linear rheology with stress fluctuation dynamics, providing a unified framework to predict and manipulate structural evolution in amorphous solids.
\end{abstract}
\maketitle
\clearpage
The structural physics and non-equilibrium thermodynamics of amorphous solids, such as metallic glasses (MGs), remain among the most intriguing puzzles in condensed matter physics~\cite{schuh2007mechanical,rodney2011modeling,Cheng2011379,sun2016thermomechanical,debenedetti2001supercooled,berthier2025yield}. Lacking long-range translational symmetry and defined crystalline defects (e.g., dislocations), MGs inherently reside in a complex rugged potential energy landscape (PEL) characterized by a multitude of local energy minima or inherent structures~\cite{debenedetti2001supercooled,heuer2008exploring,PhysRevLett.127.215502}. Under external thermomechanical perturbations, the atomic configurations undergo cooperative rearrangements within localized regions known as shear transformation zones (STZs)~\cite{argon1979plastic,Falk1998dynamics,schall2007structural}, enabling the glass to hop between different basin depths on the PEL~\cite{malandro1999relationships,Lacks2004energy,PhysRevLett.95.195501}. 
Understanding and controlling these activation and relaxation processes across energy barriers is not only of fundamental thermodynamic interest, but also dictates the mechanical performance, plasticity, and functional stability of MGs for engineering applications~\cite{greer2013shear,Wang2012a,greer2016stored}.

In recent years, cyclic mechanical deformation has emerged as a particularly versatile paradigm to navigate the PEL of amorphous solids without altering their chemical composition
~\cite{PhysRevE.88.020301,Lundberg2008reversible,priezjev2018effect,shang2022cycle,Das2022annealing,shang2025mechanical,Zhang2026,Keim2019Memory}. 
Depending on the loading parameters, most notably the strain amplitude $\gamma_A$ and temperature,
cyclic shearing can push the glass into two diametrically opposed structural trajectories: mechanical annealing (aging) or structural rejuvenation~\cite{PhysRevE.88.020301,regev2015reversibility,leishangthem2017yield}. 
Sub-yield cyclic driving typically drives the system into deeper, thermodynamically more stable energy minima, leading to structural densification, stress relaxation, and enhanced yield strength~\cite{Lundberg2008reversible,shang2022cycle,bhaumik2021role,Yeh2020glass,Khirallah2021yielding,Liu2022fate}. Conversely, over-yield cyclic deformation injects substantial plastic work, destabilizing the glassy matrix into shallower, higher-energy basins, a process termed rejuvenation that promotes ductility, shear band delocalization, and strain softening~\cite{leishangthem2017yield,Bhaumik2022avalanches,maity2026fatigue,Liang2024structural}. 

Despite extensive experimental and computational efforts, a unified, quantitative continuum framework that connects sub-yield mechanical aging to over-yield non-linear flow remains conspicuously absent. 
A primary open question centers on the fundamental connection between microscopic plastic dissipation and the kinetic pathways across energy regimes. While over-yield steady-state flow in complex fluids and glasses is traditionally described by non-linear rheological models like the Herschel-Bulkley (H-B) framework~\cite{Bonn2017yield,Nicolas2018deformation,Coussot2002coexistence,Besseling2007three,schuh2007mechanical}, 
sub-yield structural relaxation is commonly treated via empirical stretched-exponential kinetics or diffusion-like energy basin hopping~\cite{phillips1996stretched,richert2002heterogeneous,sun2023distinct}.
How these seemingly disparate dynamic regimes seamlessly interface, and whether they share a common physics rooted in driving rates ($\dot{\gamma}$), is poorly understood. Furthermore, at intermediate strain amplitudes around the nominal yield point, systems frequently display complex non-monotonic potential energy evolution—a transient crossover from initial rejuvenation to subsequent mechanical aging~\cite{leishangthem2017yield,Parmar2019strain,sun2023distinct}. The dynamic origins of this non-monotonic trajectory, the microscopic nature of the crossover period $N_c$, and the universal scaling laws governing these competing energy mechanisms demand rigorous theoretical clarification.

In this Letter, using extensive molecular dynamics (MD) simulations on a prototypical $\mathrm{Cu}_{64}\mathrm{Zr}_{36}$ glass system across a wide range of strain amplitudes ($\gamma_A$) and loading frequencies, we bridge this longstanding gap by discovering an explicit scaling connection between sub-yield aging and over-yield rheology. We uncover two distinct universal scaling exponents: an over-yield plastic flow exponent $\alpha \approx 0.17$ and a sub-yield aging exponent $\beta \approx 0.34$. By integrating local elastic stress fluctuation dynamics into the Herschel-Bulkley framework, we establish a dynamic rationale demonstrating that $\beta \approx 2\alpha$. This result reveals that while over-yield rejuvenation represents a first-order direct response to macroscopic plastic shear work ($\propto \dot{\gamma}^\alpha$), sub-yield mechanical aging is mediated by a second-order local elastic stress fluctuation mechanism ($\langle \delta\sigma_{\mathrm{local}}^2 \rangle \propto \dot{\gamma}^{2\alpha}$). Finally, by framing the characteristic crossover cycle $N_c$ as a dynamical balance between plastic work injection and elastic fluctuation dissipation, we demonstrate a universal master collapse of the crossover boundaries across dynamic regimes, providing a predictive protocol for tailoring the structural state of amorphous materials via cyclic mechanical training.


To probe the structural evolution under cyclic mechanical drive, classical MD simulations were conducted using the LAMMPS package~\cite{Plimpton1995fast} on a model binary glass, $\mathrm{Cu}_{64}\mathrm{Zr}_{36}$, comprising $16,000$ atoms. Interatomic interactions were modeled via a well-validated embedded-atom method (EAM) potential~\cite{mendelev2009develop}. Periodic boundary conditions were enforced in all directions, and the equations of motion were integrated with a time step of $2~\mathrm{fs}$ within the isothermal-isobaric ($NPT$) ensemble at zero external pressure. Initial glass configurations were prepared by equilibrating the liquid phase at $2000~\mathrm{K}$, followed by quenching to $300~\mathrm{K}$ at a constant cooling rate of $1~\mathrm{K/ps}$. Prior to deformation, the quenched samples were further relaxed at $300~\mathrm{K}$ for $1~\mathrm{ns}$ to establish thermal stability.
Cyclic uniaxial tension-compression was applied along the $x$-direction under a sinusoidal strain protocol, $\gamma(t) = \gamma_A \sin(2\pi t / t_p)$, where $\gamma_A$ represents the strain amplitude and $t_p$ is the loading period (spanning $t_p = 10\text{--}1000~\mathrm{ps}$).
The characteristic strain rate is thus defined as $\dot{\gamma} \equiv \gamma_A / t_p$. 
A representative cyclic stress and energy response at $\gamma_A = 0.06$ is illustrated in Figs.~\ref{fig:S1} and \ref{fig:S2} in the Supplemental Material, displaying characteristic periodic fluctuations throughout the loading history. 
To eliminate transient elastic strain energy, the system's potential energy was recorded strictly at the zero-strain states ($\gamma = 0$) at the end of each loading cycle. Statistical ensemble averages were computed over $10$ independent configurations prepared via the isoconfigurational method.

\begin{figure}[htbp]
\centering
\includegraphics[width=0.9\linewidth]{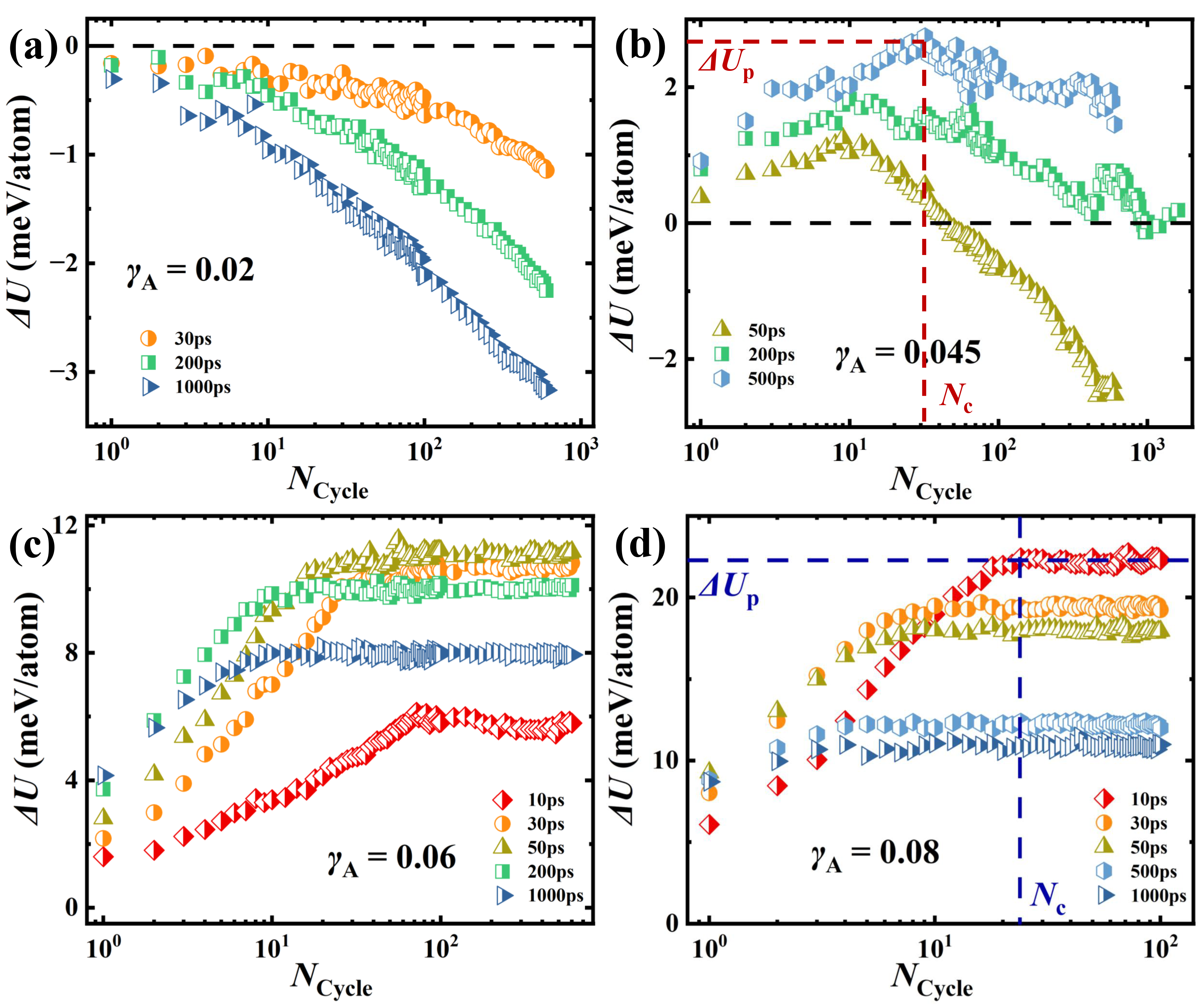}
\caption{Representative evolution of potential energy change, $\Delta U$, versus the number of loading cycles, $N_{\mathrm{cycle}}$, under different strain amplitudes and frequencies in $\mathrm{Cu}_{64}\mathrm{Zr}_{36}$. (a) Monotonic energy decrease (mechanically accelerated aging) at $\gamma_A=0.02$. (b) Initial increase to a maximum value $\Delta U_p$ followed by a decrease at $\gamma_A=0.045$; the horizontal dashed lines in (a,b) mark $\Delta U=0$. (c,d) Increase toward a dynamic steady state with a plateau at $\gamma_A=0.06$ and $0.08$, respectively. The plateau value or peak value in (b) is denoted as $\Delta U_p$, and the corresponding crossover cycle number is $N_c$, as illustrated by the dashed line in (d).}
\label{fig:1}
\end{figure}

To map the dynamical response of the system under cyclic deformation, we track the evolution of potential energy change, $\Delta U$, as a function of loading cycle number, $N_{\mathrm{cycle}}$. Figure~\ref{fig:1} displays four representative response regimes across different strain amplitudes $\gamma_A$.
At small strain amplitudes ($\gamma_A = 0.02$, Fig.~\ref{fig:1}(a)), the system exhibits monotonic energy dissipation below its initial un-deformed state ($\Delta U < 0$), signaling mechanically accelerated aging. Lower loading frequencies (longer loading periods $t_p$) noticeably enhance the relaxation rate without reaching a steady-state plateau within $100$ cycles.
At an intermediate amplitude ($\gamma_A = 0.045$, Fig.~\ref{fig:1}(b)), $\Delta U$ exhibits a non-monotonic trajectory, initially increasing to a peak value $\Delta U_p$ before dropping, revealing a transient crossover between rejuvenation and aging. 
At higher strain amplitudes ($\gamma_A = 0.06$ and $0.08$, Figs.~\ref{fig:1}(c,d)), the system undergoes sustained rejuvenation, monotonically increasing toward a steady-state energy platform $\Delta U_p$ at a characteristic crossover cycle number $N_c$.

\begin{figure}[htbp]
\centering
\includegraphics[width=0.9\linewidth]{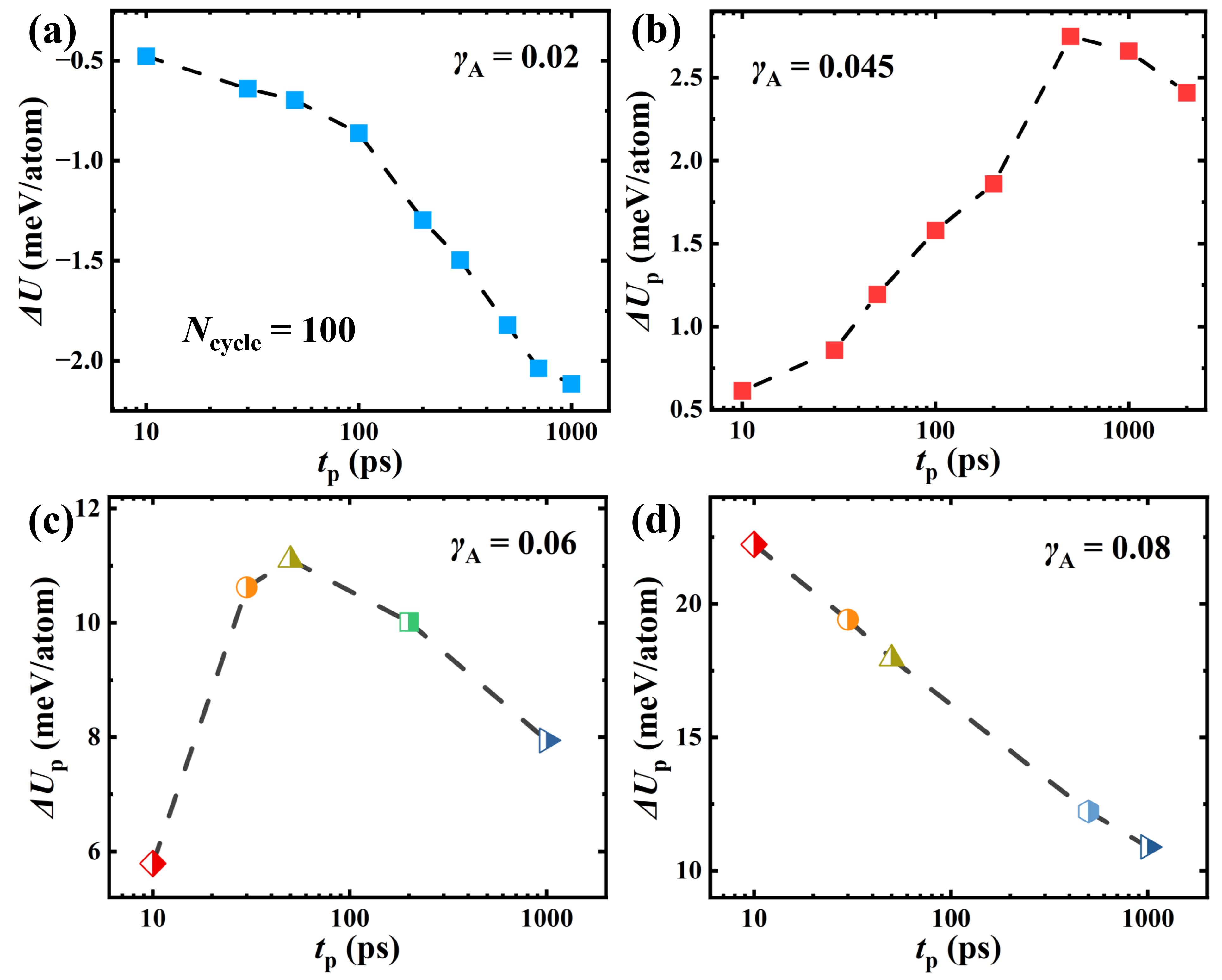}
\caption{Frequency dependence of the characteristic potential energy for the four representative strain amplitudes $\gamma_A$ shown in Fig.~\ref{fig:1}. (a) $\Delta U$ at the 100th loading cycle as a function of loading period $t_p$ for $\gamma_A=0.02$. (b) Maximum potential energy change $\Delta U_p$ as a function of $t_p$ for $\gamma_A=0.045$. (c,d) Steady-state plateau energy $\Delta U_p$ as a function of $t_p$ for $\gamma_A=0.06$ and $0.08$, respectively.}
\label{fig:2}
\end{figure}

The rate dependency of these characteristic energy states ($\Delta U$ or $\Delta U_p$) demonstrates a dramatic dynamic reversal across amplitudes, as summarized in Fig.~\ref{fig:2}. In the sub-yield aging regime ($\gamma_A = 0.02$, Fig.~\ref{fig:2}(a)), the energy state evaluated at $N_{\mathrm{cycle}}=100$ decreases monotonically with increasing loading period $t_p$, proving that lower strain rates allow more time for atomic exploration into deeper energy basins. Surprisingly, at intermediate amplitudes ($\gamma_A = 0.045$ and $0.06$, Figs.~\ref{fig:2}(b,c) ), $\Delta U_p$ exhibits a non-monotonic dependence on $t_p$, rising initially before dropping at longer periods (see also Fig.~\ref{fig:S3} in the Supplemental Material). Upon further increasing the amplitude into the fully over-yield flow regime ($\gamma_A = 0.08$, Fig.~\ref{fig:2}(d)), the rate dependence completely reverses: $\Delta U_p$ decreases monotonically with $t_p$.

\begin{figure}[htbp]
\centering
\includegraphics[width=0.9\linewidth]{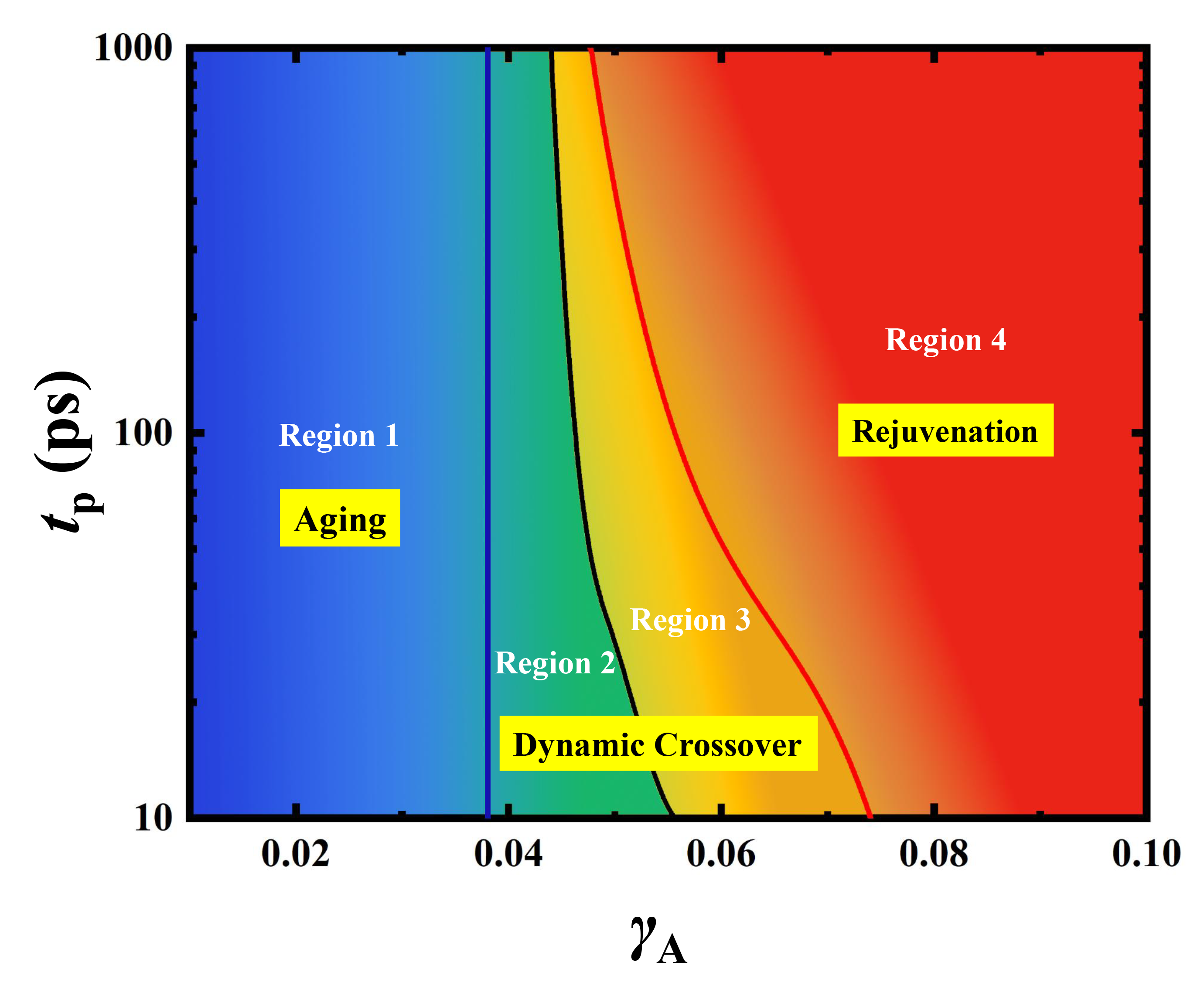}
\caption{Dynamic phase diagram of cyclically driven $\mathrm{Cu}_{64}\mathrm{Zr}_{36}$ metallic glass in the strain amplitude--frequency plane. The cyclic response evolves from mechanically accelerated aging (Region 1), through a dynamic crossover between rejuvenation and aging (Regions 2 and 3), to sustained rejuvenation (Region 4). The frequency axis is plotted on a logarithmic scale.}
\label{fig:3}
\end{figure}

Synthesizing these kinetic behaviors across the complete $(\gamma_A, t_p)$ parameter space yields the dynamical phase diagram shown in Fig.~\ref{fig:3}. The state space is partitioned into four operational domains: Region 1 (purely mechanically accelerated aging), Regions 2 and 3 (a dynamic crossover zone characterized by competitive rejuvenation/aging kinetics), and Region 4 (sustained over-yield mechanical rejuvenation).

\begin{figure}[htbp]
\centering
\includegraphics[width=0.9\linewidth]{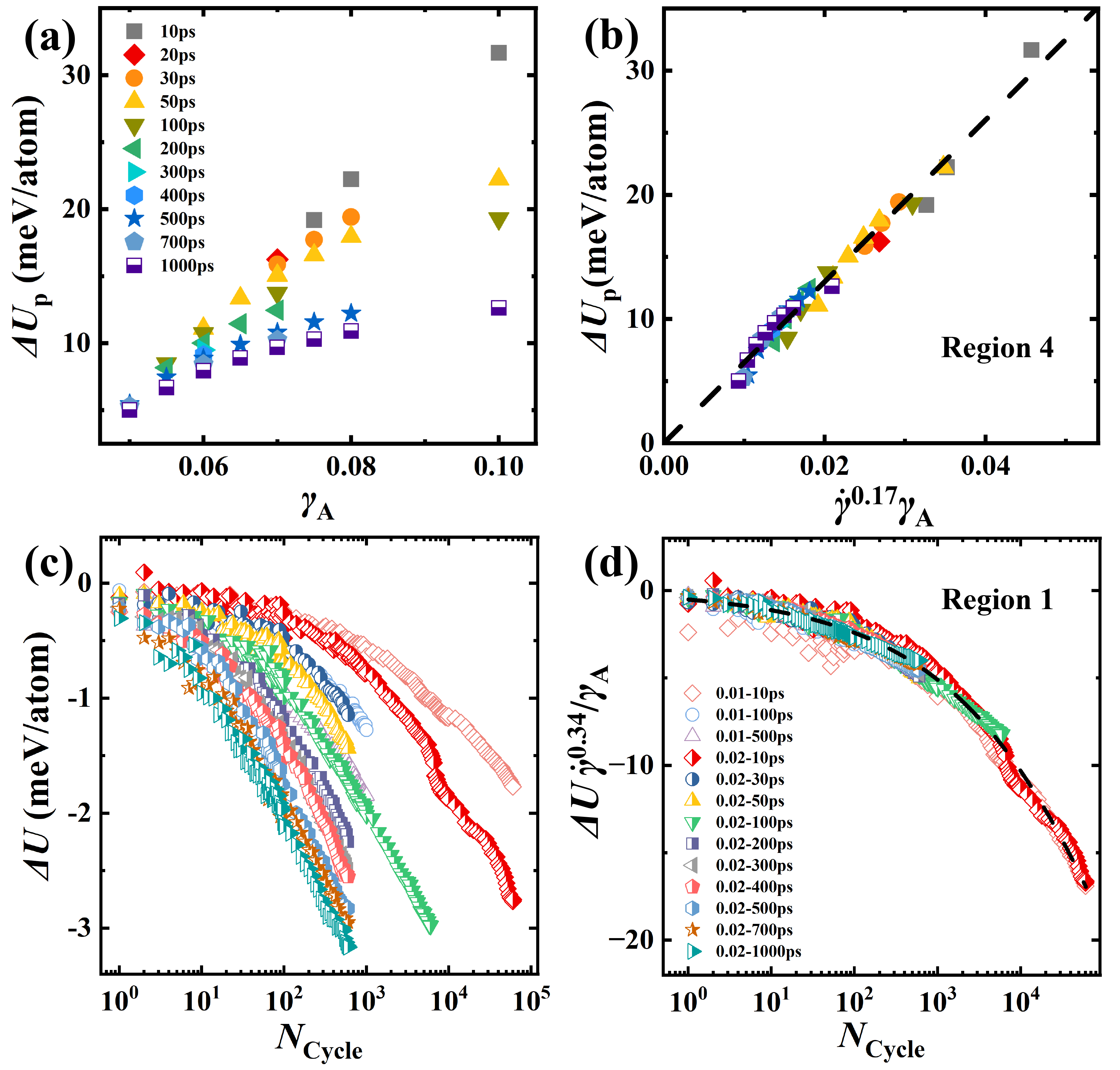}
\caption{Universal scaling of mechanically driven rejuvenation and aging in $\mathrm{Cu}_{64}\mathrm{Zr}_{36}$. (a) Plateau energy $\Delta U_p$ as a function of $\gamma_A$ in Region 4. (b) Rheological scaling collapse showing $\Delta U_p \propto \gamma_A \dot{\gamma}^{0.17}$ based on the Herschel-Bulkley model; the dashed line is a linear fit. (c) Evolution of $\Delta U$ versus $N_{\mathrm{cycle}}$ for various frequencies at small $\gamma_A$. (d) Master curve collapse achieved by scaling energy as $\Delta U \dot{\gamma}^{0.34}/\gamma_A$, consistent with the proposed relation $\beta \approx 2\alpha = 0.34$. The dashed line denotes a stretched-exponential fit.}
\label{fig:4}
\end{figure}

To elucidate the microscopic mechanisms governing these operational domains, we analyze the scaling behaviors in Regions 1 and 4 (Fig.~\ref{fig:4}). In the sustained rejuvenation regime (Region 4), the system operates above yield, where non-linear plastic flow dominates. The macroscopic flow stress obeys the classic Herschel-Bulkley (H-B) rheological relation:
\begin{equation}
\sigma = \sigma_y + K \dot{\gamma}^\alpha,
\label{eq:HB}
\end{equation}
where $\sigma_y$ is the dynamic yield stress, $K$ is the consistency index, and $\alpha$ is the shear-thinning exponent. The steady-state rejuvenation energy platform $\Delta U_p$, which scales with the macroscopic plastic work per cycle ($\propto \sigma \cdot \gamma_A$)\cite{shang2025mechanical}, scales nonlinearly with strain amplitude $\gamma_A$ (Fig.~\ref{fig:4}(a)). As shown in Fig.~\ref{fig:4}(b), all plateau energy values collapse onto a single linear master curve when plotted against $\gamma_A \dot{\gamma}^\alpha$:
\begin{equation}
\Delta U_p \propto \gamma_A \dot{\gamma}^\alpha \quad (\text{with } \alpha \approx 0.17).
\label{eq:rejuvenation}
\end{equation}

In contrast, for the sub-yield aging regime (Region 1, Fig.~\ref{fig:4}(c)), scaling the potential energy evolution by $\dot{\gamma}^\beta / \gamma_A$ collapses all aging curves across various frequencies onto a single master curve (Fig.~\ref{fig:4}(d)):
\begin{equation}
\frac{\Delta U}{\gamma_A} \dot{\gamma}^\beta = \mathcal{F}(N_{\mathrm{cycle}}) \quad (\text{with } \beta \approx 0.34),
\label{eq:aging}
\end{equation}
where $\mathcal{F}(N_{\mathrm{cycle}})$ is accurately fitted by a stretched-exponential decay function.

Remarkably, the sub-yield aging exponent ($\beta = 0.34$) is nearly twice the over-yield flow exponent ($\alpha = 0.17$), i.e., $\beta \approx 2\alpha$. This scaling relationship suggests a plausible physical origin connecting sub-yield relaxation to over-yield rheology within the H-B framework. In the sub-yield regime ($\gamma_A < \gamma_y$), global plastic flow is suppressed; however, structural heterogeneity gives rise to spatial fluctuations in local elastic stress, $\delta \sigma_{\mathrm{local}}$\cite{barrat2011heterogeneities,Nicolas2018deformation,PhysRevLett.126.138005}. 
Assuming these local stress fluctuations inherit the characteristic stress scale of the driven matrix ($\delta \sigma_{\mathrm{local}} \propto \Delta \sigma_{\mathrm{macro}} \propto \dot{\gamma}^\alpha$), the effective mechanical agitation energy per cycle can be estimated from second-order elastic strain energy fluctuations~\cite{Tsamados2009local}:
\begin{equation}
\Delta E_{\mathrm{agitation}} \propto \frac{\langle \delta \sigma_{\mathrm{local}}^2 \rangle}{2G} \propto \left( \dot{\gamma}^\alpha \right)^2 = \dot{\gamma}^{2\alpha},
\label{eq:fluctuation}
\end{equation}
where $G$ is the effective shear modulus. In the sub-yield regime, this local agitation functions as an effective mechanical noise/temperature that destabilizes potential energy basins~\cite{Falk1998dynamics, sollich1997rheology,Falk2011deformation}. Rapid agitation increases the escape probability from deeper potential minima~\cite{Lacks2004energy}, rendering the net retention efficiency of structural relaxation inversely proportional to the agitation intensity ($\text{Efficiency} \propto 1 / \Delta E_{\mathrm{agitation}} \propto \dot{\gamma}^{-2\alpha}$). Scaling the single-cycle potential energy reduction with this relaxation retention efficiency ($\Delta U / \gamma_A \propto \text{Efficiency} \propto \dot{\gamma}^{-\beta}$) directly connects the potential energy drop to local stress fluctuations, naturally yielding the identity:
$\beta \approx 2\alpha = 0.34$.

This microscopic mechanism physically conveys that intense mechanical agitation at elevated strain rates prevents the atomistic system from remaining trapped in deeper energy wells, suppressing the net potential energy drop per strain cycle.
While these quantitative exponent values are extracted from our representative $\mathrm{Cu}_{64}\mathrm{Zr}_{36}$ model system, this fluctuation-based scaling rationale offers a transparent physical picture: over-yield rejuvenation is predominantly driven by first-order macroscopic plastic shear work, whereas sub-yield aging appears to be mediated by second-order local elastic stress fluctuations.

\begin{figure}[htbp]
\centering
\includegraphics[width=0.9\linewidth]{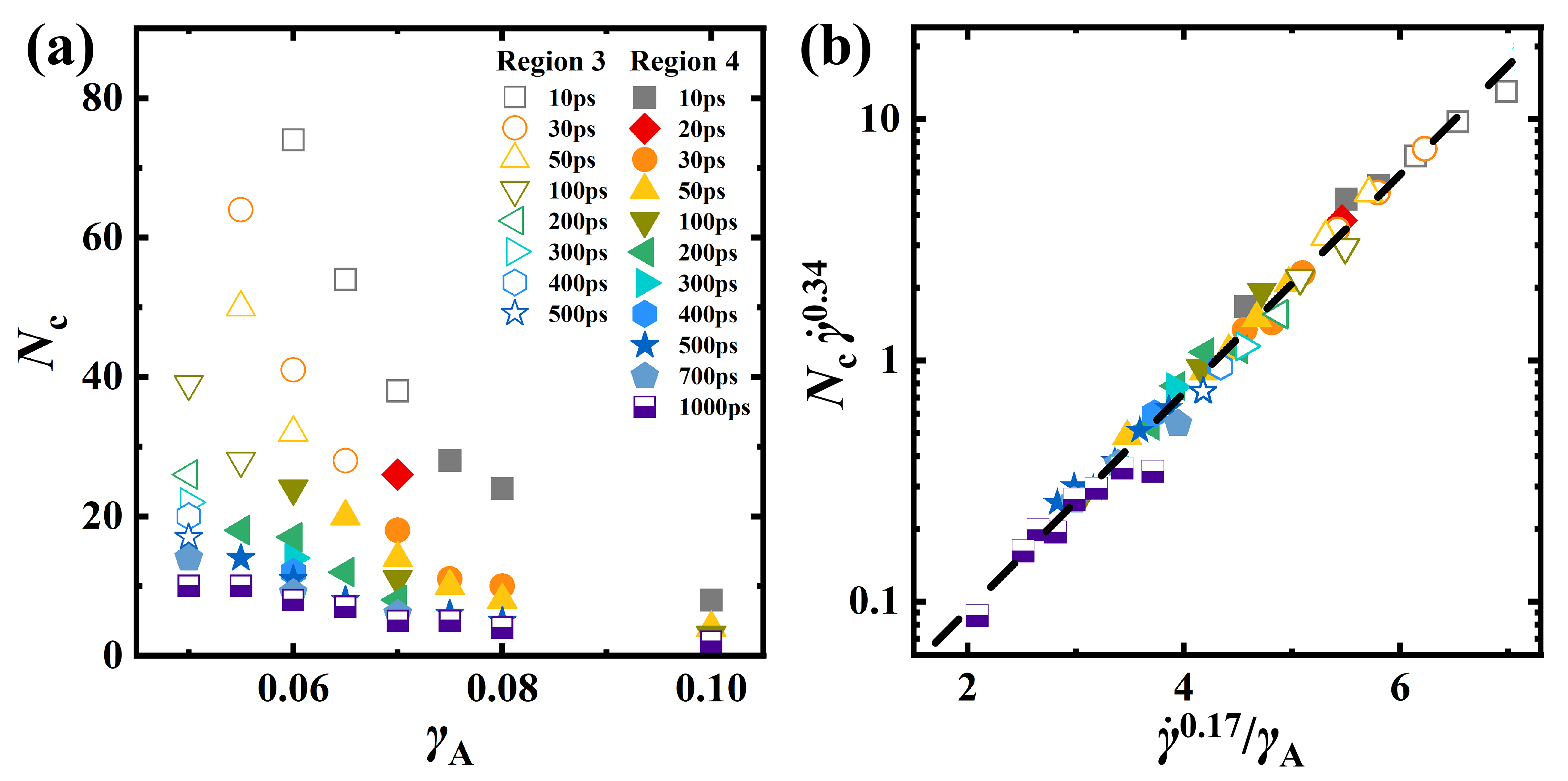}
\caption{Universal scaling of rejuvenation dynamics and the kinetic competition between energy mechanisms. (a) Characteristic cycle number $N_c$ required to reach $\Delta U_p$ as a function of $\gamma_A$. (b) Master scaling collapse of $N_c \dot{\gamma}^{0.34}$ against $\dot{\gamma}^{0.17}/\gamma_A$ for Regions 3 and 4, validating the universal function $N_c \dot{\gamma}^{2\alpha} = \mathcal{G}(\dot{\gamma}^\alpha/\gamma_A)$.}
\label{fig:5}
\end{figure}

Finally, we address the dynamic crossover in Regions 2 and 3, where transient rejuvenation and aging compete directly (Fig.~\ref{fig:5}). The characteristic cycle number $N_c$ required to reach the peak or plateau energy $\Delta U_p$ decreases rapidly with increasing strain amplitude (Fig.~\ref{fig:5}(a)). 

The observed transition from early-stage rejuvenation to long-term aging reflects a fundamental separation in time scales governing the elementary deformation processes~\cite{Lacks2004energy,leishangthem2017yield}. Short-term cyclic shear primarily acts as a direct excitation driver ($N < N_c$), promoting localized atomic jumps over potential energy barriers into higher-energy inherent structures.
Conversely, long-term structural relaxation is driven by second-order stress fluctuations ($\langle\delta\sigma^2\rangle$) generated by these discrete plastic events~\cite{Nicolas2018deformation}. These quadrupolar elastic interactions require cumulative loading cycles ($N > N_c$) to establish an effective structural agitation field capable of driving the system into deeper energy basins\cite{Khirallah2021yielding}. This mechanistic decoupling produces a temporal lag between local excitation and collective relaxation, directly accounting for the non-monotonic trajectory and the dynamic crossover observed at $N_c$.

Coupling the cumulative magnitude of second-order elastic agitation ($N_c \dot{\gamma}^{2\alpha}$) with the relative plastic-to-elastic energy ratio ($\propto \dot{\gamma}^\alpha \gamma_A / \gamma_A^2 = \dot{\gamma}^\alpha / \gamma_A$) yields a general scaling relation across Regions 3 and 4:
\begin{equation}
N_c \dot{\gamma}^{2\alpha} = \mathcal{G}\left( \frac{\dot{\gamma}^\alpha}{\gamma_A} \right),
\label{eq:master_collapse}
\end{equation}
where $\mathcal{G}$ is a universal scaling function (Fig.~\ref{fig:5}(b)).

It is essential to clarify the physical boundaries and validity of this scaling collapse across different operational domains. For Region~3 and Region~4, an energy-normalized representation based on the peak structural potential energy elevation, $\Delta U_p / (\frac{1}{2}E\gamma_A^2 V_0)$, collapses Region~4 data onto the predicted scaling line (${\Delta U_p}/{W_{\text{elastic}}} \sim N_c \dot{\gamma}^{2\alpha}$), but systematic upward deviations occur in Region~3 (and Region~2) [Sec.~3 and Fig.~\ref{fig:S5} in Supplemental Material]. This deviation arises because, during the transient transition, local structural rearrangements allow the system to accumulate potential energy faster per cycle than predicted by uniform rate scaling~\cite{Falk1998dynamics}. Replacing the accumulated peak potential energy $\Delta U_p$ with the single-cycle effective external work ratio---defined via the Herschel-Bulkley flow over-stress as ${W_{\text{ext}}}/{W_{\text{elastic}}} \propto {\sigma_{\text{flow}}\gamma_A}/{\gamma_A^2} = {\dot{\gamma}^\alpha}/{\gamma_A}$---cleanly strips away transient non-structural viscous dissipation. This isolates the dynamic driving force required for structural elevation, achieving the universal master collapse for both Region~3 and Region~4 in Fig.~\ref{fig:5}(b).


For Region~2, the scaling framework in Eq.~(\ref{eq:master_collapse}) naturally breaks down, with data points systematically falling below the master curve and overestimating the required cycles to reach the potential energy peak ($N_c^{\text{pred}} > N_c^{\text{MD}}$). In this sub-yield regime ($\gamma_A < \gamma_y$), the macroscopic first-order flow work ($W_{\text{ext}} \propto \dot{\gamma}^\alpha$) is extremely small and insufficient to account for the accelerated early-stage structural elevation~\cite{Shohat2026aging}. Instead, structural evolution in Region~2 is initially governed by second-order localized elastic stress fluctuations ($\langle \delta\sigma_{\text{local}}^2 \rangle \propto \dot{\gamma}^{2\alpha} \approx \dot{\gamma}^{0.34}$)~\cite{Nicolas2018deformation}. 

During the initial cycling phase, these intense local stress fluctuations act as a ``mechanical catalyst'' that lowers the activation barriers for shear transformation zones (STZs)~\cite{argon1979plastic,rodney2011modeling,argon2013strain}, enabling the system to reach its peak potential energy $\Delta U_p$ in far fewer cycles than predicted by macroscopic work. Beyond this critical peak $N_c$, the system enters a long-term aging regime. Here, these same localized fluctuations serve as a mechanical noise source that increases the de-trapping probability from deep potential basins, suppressing the net retention efficiency of deep structural relaxation ($\Delta U / \gamma_A \propto \dot{\gamma}^{-\beta}$) and directly yielding the identity $\beta \approx 2\alpha = 0.34$.

This scaling collapse highlights that $N_c \dot{\gamma}^{2\alpha}$ acts as a dimensionless intrinsic clock dictating the elastic agitation needed to navigate the energy landscape~\cite{schuh2007mechanical}. Exceeding $N_c$ triggers persistent stress fluctuations that drive steady-state flow or long-term structural relaxation~\cite{Shohat2026aging}.

It is worth noting that the exponents $\alpha$ and $\beta$ depend on the deformation path and boundary conditions. Under uniaxial $NVT$ tension, Sun et al.~\cite{sun2023distinct} obtained $\beta = 3/7 \approx 0.43$. Their lateral confinement induces hydrostatic tension, raising the volumetric activation barrier for shear transformation zones (STZs) and increasing rate sensitivity. Similarly, Shang et al.~\cite{shang2025mechanical} observed $\alpha \approx 0.50$ under triaxial cyclic deformation , where multiaxial constraints impede single-plane shear localization.

In contrast, our uniaxial $NPT$ tensile simulations allow unconstrained lateral Poisson contraction. This facilitates local free-volume redistribution, leading to a milder rate dependence ($\alpha \approx 0.17, \beta \approx 0.34$). Crucially, while individual values of $\alpha$ and $\beta$ vary with mechanical constraints, the fluctuation-driven framework implies an intrinsic connection between stress fluctuations and relaxation dynamics ($\beta \approx 2\alpha$). A systematic verification of this scaling relation across diverse loading modes and stress states remains a promising direction for future in-depth investigation.

In summary, using a representative $\text{Cu}_{64}\text{Zr}_{36}$ model system, we have developed a unified thermodynamic-rheological framework for cyclically deformed metallic glasses. By combining the Herschel-Bulkley model with local elastic stress fluctuations, we provide a physical interpretation for the observed relation $\beta \approx 2\alpha$ ($\beta = 0.34$, $\alpha = 0.17$). Furthermore, framing $N_c$ through the dynamic competition between first-order macro-rheological work input and second-order local stress fluctuation assistance offers a predictive framework for navigating the energy landscape in amorphous solids via mechanical training~\cite{sun2016thermomechanical,schuh2007mechanical,Keim2019Memory}.

\begin{acknowledgments}
This work was supported by the National Natural Science Foundation of China (Grant No.12474187, No.52201169), Hebei Natural Science Foundation (Grant No.E2026202283).
\end{acknowledgments}

%

\clearpage
\pagebreak

\setcounter{equation}{0}
\setcounter{figure}{0}
\setcounter{table}{0}
\setcounter{section}{0}
\setcounter{page}{1}

\makeatletter
\renewcommand{\theequation}{S\arabic{equation}}
\renewcommand{\thefigure}{S\arabic{figure}}
\renewcommand{\thesection}{S\arabic{section}}
\renewcommand{\thepage}{S\arabic{page}}

\begin{center}
{\Large \bf Supplemental Material for:\\ ``Universal Scaling and Fluctuation-Driven Dynamic Reversal in a Cyclically Driven Metallic Glass''}\\[0.3cm]

Nannan Ren,$^1$ Hongfei Zhang,$^1$ Rui Wan,$^1$ Pengfei Guan,$^{2,*}$ Limin Wang,$^{3,\dagger}$ and Baoshuang Shang$^{3,\ddagger}$\\[0.1cm]

{\small $^1${\it School of Materials Science and Engineering, Anhui University of Technology, Maanshan 243032, China}}\\
{\small $^2${\it Advanced Interdisciplinary Science Research Center (AiRCenter), Ningbo Institute of Materials Technology and Engineering, Chinese Academy of Sciences, Ningbo 315201, China}}\\
{\small $^3${\it State Key Laboratory of High Performance Roll Materials and Composite Forming, School of Material Science and Engineering, Hebei University of Technology, Tianjin 300401, China}}\\[0.1cm]

{\small $^*$Electronic address: pguan@nimte.ac.cn}\\
{\small $^\dagger$Electronic address: Liminwang@hebut.edu.cn}\\
{\small $^\ddagger$Electronic address: shangbaoshuang@hebut.edu.cn}\\[0.5cm]
\end{center}
\subsection*{1. Cyclic loading protocol and representative deformation response}

\begin{figure}[htbp]
\centering
\includegraphics[width=0.8\linewidth]{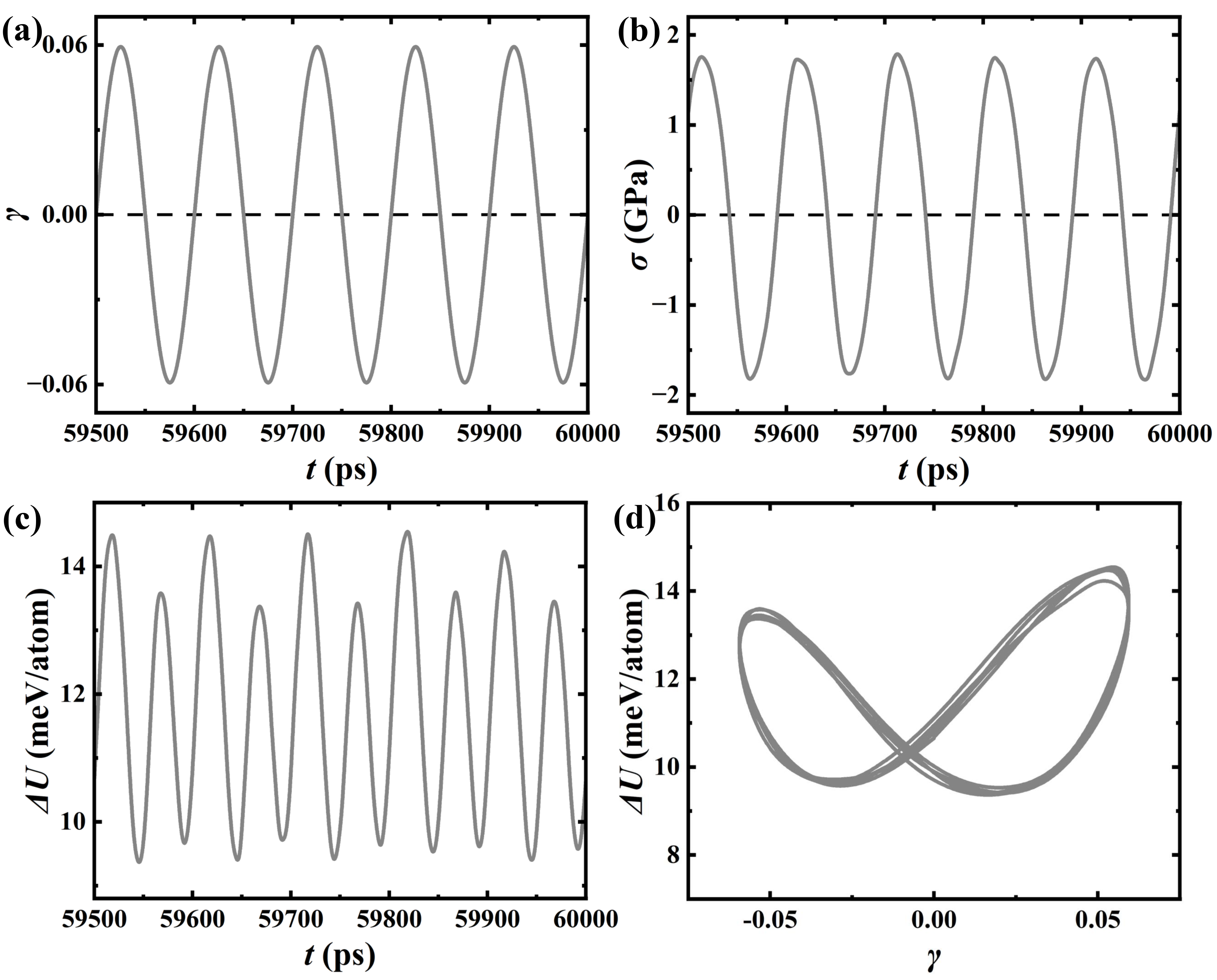}
\caption{Representative cyclic deformation responses during the final five loading cycles at a strain amplitude of $\gamma_A=0.06$. (a) Applied strain $\gamma$, (b) stress $\sigma$ and (c) potential energy change, $\Delta U$, as a function of time. (d) $\Delta U$ as a function of strain.}
\label{fig:S1}
\end{figure}

\begin{figure}[htbp]
\centering
\includegraphics[width=0.8\linewidth]{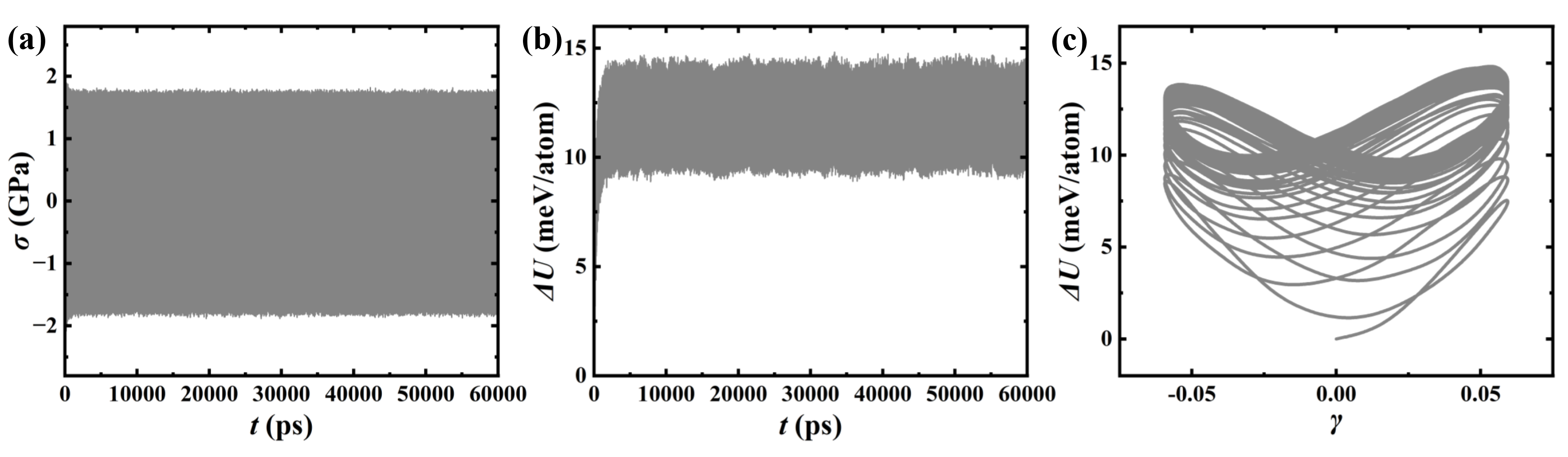}
\caption{Representative cyclic deformation responses at $\gamma_A=0.06$. (a) Stress and (b) potential energy as functions of time, and (c) $\Delta U$ as a function of strain throughout the entire cyclic loading process.}
\label{fig:S2}
\end{figure}

To illustrate the cyclic loading protocol, Fig.~\ref{fig:S1} shows the representative responses during the final five loading cycles at $\gamma_A=0.06$. The imposed strain follows a periodic waveform [Fig.~\ref{fig:S1}(a)], resulting in a corresponding cyclic stress response [Fig.~\ref{fig:S1}(b)]. The potential energy change $\Delta U$ exhibits periodic evolution with time [Fig.~\ref{fig:S1}(c)] and forms a characteristic energy-strain loop during cyclic loading [Fig.~\ref{fig:S1}(d)]. These results illustrate the deformation protocol used throughout the simulations. Besides, Fig.~\ref{fig:S2} shows the representative responses during the entire cyclic loading process at $\gamma_A=0.06$. The stress and potential energy exhibit periodic variations with time, while the potential energy change $\Delta U$ evolves systematically with the applied strain. These results illustrate the cyclic loading protocol and the corresponding mechanical and energetic responses used in the simulations.

\begin{figure}[htbp]
\centering
\includegraphics[width=0.8\linewidth]{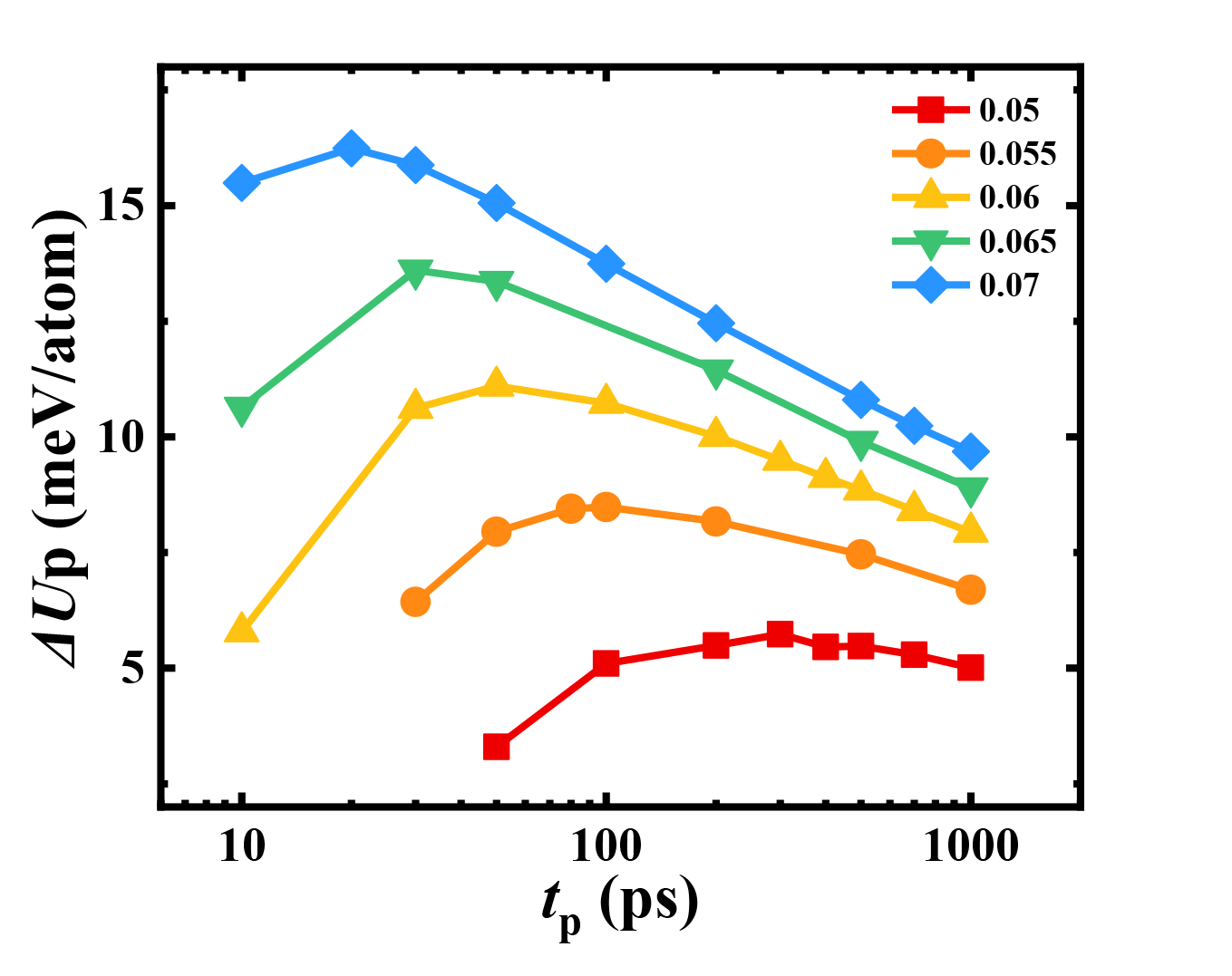}
\caption{$\Delta U_p$ as a function of loading frequency for different strain amplitudes $\gamma_A$.}
\label{fig:S3}
\end{figure}

\begin{figure}[htbp]
\centering
\includegraphics[width=0.8\linewidth]{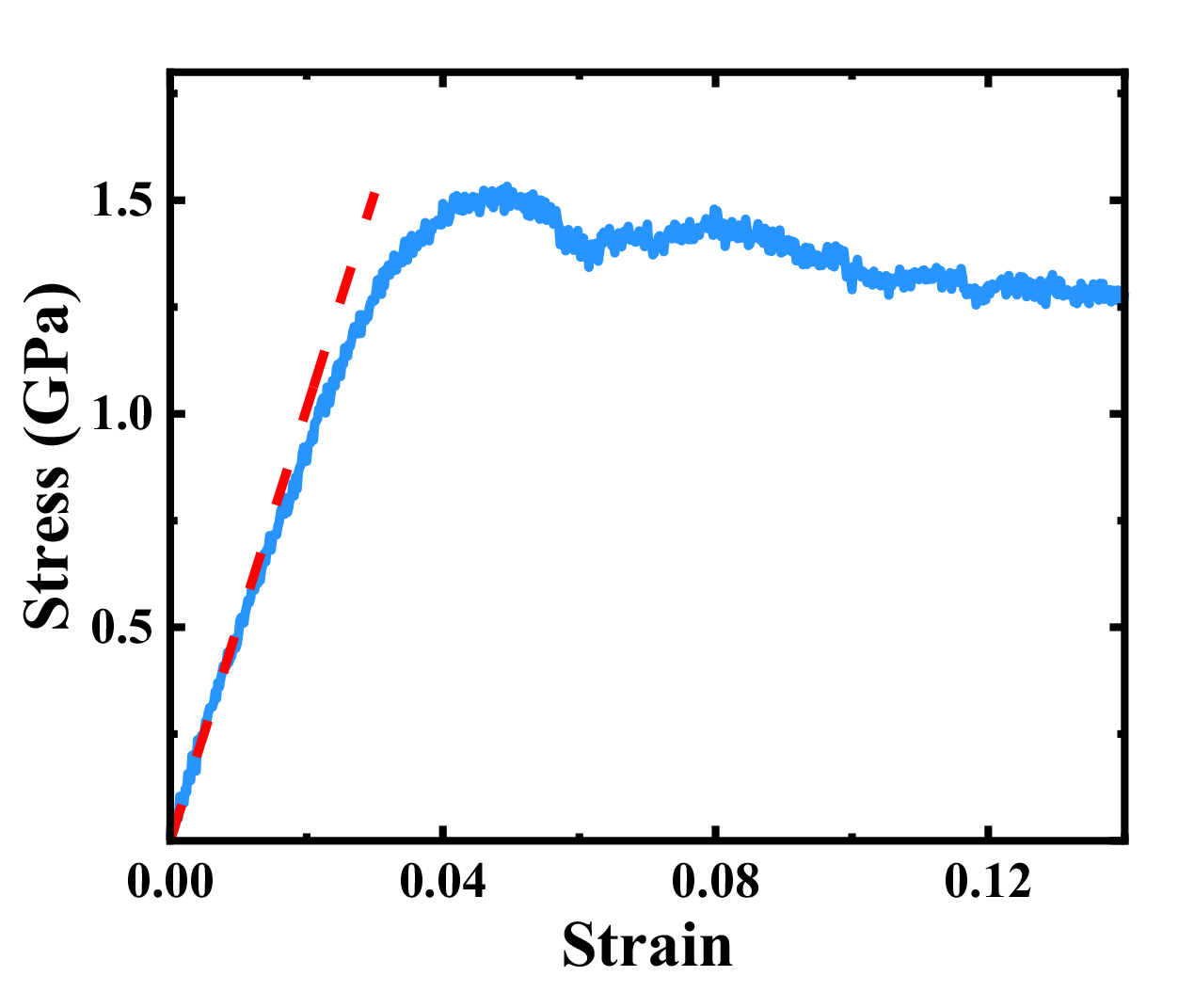}
\caption{Stress--strain response of the as-prepared MG sample at a strain rate of $1\times10^7~\mathrm{s}^{-1}$. The dashed line represents a linear fit to the small-strain region, yielding an elastic modulus of $E=50.6~\mathrm{GPa}$.}
\label{fig:S4}
\end{figure}

\subsection*{2. Stretched-exponential formulation of the sub-yield aging master curve (Region 1)}

In the sub-yield regime (Region 1, small strain amplitudes), the potential energy evolution $\Delta U(N_{\mathrm{cycle}})$ under cyclic mechanical drive collapses onto a universal master curve when scaled by $\dot{\gamma}^{0.34} / \gamma_A$, as shown in Fig.~4(d) of the main text. This structural relaxation process is characterized by hierarchical energetic barrier crossing, which is quantitatively described by a stretched-exponential decay function:
\begin{equation}
\frac{\Delta U}{\gamma_A} \dot{\gamma}^{0.34} = A \exp \left[ -\left( \frac{N_{\mathrm{cycle}}}{N_0} \right)^\beta \right]-A,
\label{eq:S_stretched_exp}
\end{equation}
where $A$ is the normalized energy amplitude, $N_0$ is the characteristic cycle number for relaxation, and $\beta$ is the stretching exponent ($0 < \beta \le 1$). This stretched-exponential form explicitly underscores the cooperative and heterogeneous nature of localized atomic movements driven by second-order elastic stress fluctuations ($\propto \dot{\gamma}^{2\alpha} = \dot{\gamma}^{0.34}$) under sub-yield cyclic deformation.

\subsection*{3. Dynamics of approach to the steady state, mechanism decoupling, and boundaries of scaling collapses}
\begin{figure}[htbp]
\centering
\includegraphics[width=0.9\linewidth]{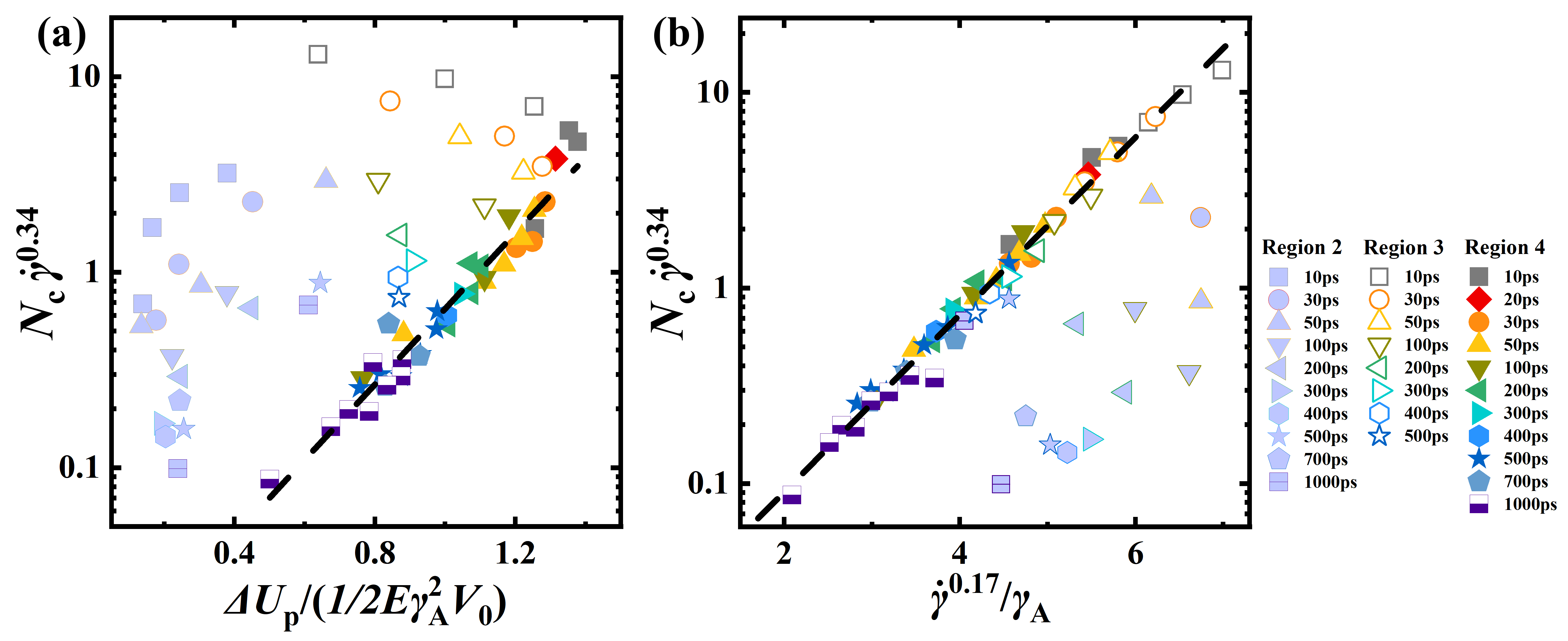}
\caption{Scaling collapse of the characteristic crossover cycle $N_c$ across operational domains. (a) Energy-normalized representation $N_c \dot{\gamma}^{0.34}$ as a function of $\Delta U_p / (\frac{1}{2}E\gamma_A^2V_0)$ for Regions 2, 3, and 4. The dash line represents the exponential fit [Eq.~(\ref{eq:S_energy_collapse})] describing Region 4 data, with systematic upward deviations occurring in Regions 2 and 3 due to transient structural rearrangement dynamics. (b) Universal master collapse of Regions 3 and 4 achieved by parameterizing against the single-cycle effective work ratio $\dot{\gamma}^{0.17}/\gamma_A$ [Eq.~(\ref{eq:S_rate_collapse})]. Region 2 data points lie below the master curve due to acceleration by second-order localized elastic stress fluctuations.}
\label{fig:S5}
\end{figure}
To further delineate the rejuvenation dynamics across different operational domains, we first evaluate an energy-based scaling representation by normalizing the peak/plateau potential energy change $\Delta U_p$ by the macroscopic elastic strain energy scale, $\frac{1}{2}E\gamma_A^2V_0$. Here, $E = 50.6~\mathrm{GPa}$ is the elastic modulus extracted from the linear regime of the stress--strain curve (Fig.~\ref{fig:S4}), and $V_0$ is the initial atomic volume of the sample.

As shown in Fig.~\ref{fig:S5}(a), an energy-normalized scaling representation based on the peak structural potential energy elevation, $\Delta U_p / (\frac{1}{2}E\gamma_A^2 V_0)$, collapses the data in Region~4 (sustained over-yield rejuvenation) onto an exponential scaling curve:
\begin{equation}
N_c \dot{\gamma}^{0.34} = 0.0076 \exp \left[ 4.46 \frac{\Delta U_p}{\frac{1}{2}E\gamma_A^2 V_0} \right].
\label{eq:S_energy_collapse}
\end{equation}
However, systematic upward deviations occur in Regions~2 and 3. This indicates that during these transient transitions, rapid local structural rearrangements allow the system to elevate its potential energy more efficiently per cycle than predicted by uniform steady-state scaling.

To eliminate non-structural viscous dissipation and isolate the intrinsic driving force for structural elevation across all over-yield regimes, the accumulated peak energy $\Delta U_p$ is replaced by the single-cycle effective external work ratio, ${W_{\text{ext}}}/{W_{\text{elastic}}} \propto {\sigma_{\text{flow}}\gamma_A}/{\gamma_A^2} = {\dot{\gamma}^{0.17}}/{\gamma_A}$, derived from the Herschel-Bulkley flow over-stress. As depicted in Fig.~\ref{fig:S5}(b) [and Main Text Fig.~5(b)], this reformulation collapses both Region~3 and Region~4 universally onto the master scaling function $\mathcal{G}(\dot{\gamma}^\alpha / \gamma_A)$:
\begin{equation}
N_c \dot{\gamma}^{0.34} = 0.011 \exp \left( 1.04 \frac{\dot{\gamma}^{0.17}}{\gamma_A} \right).
\label{eq:S_rate_collapse}
\end{equation}

A comprehensive summary of the scaling behaviors across all operational domains is detailed below:

\begin{enumerate}
    \item \textbf{Potential Energy Peak Scaling} [Fig.~\ref{fig:S5}(a), Eq.~(\ref{eq:S_energy_collapse})]: \\
    When parameterized by the maximum potential energy elevation $\Delta U_p$, data points for Regions~2 and 3 systematically lie \textit{above} the scaling line [Eq.~(\ref{eq:S_energy_collapse})]. This demonstrates enhanced structural potential energy storage per cycle during the transient phase. Only in Region~4, where the system reaches a dynamic steady-state saturation plateau, do the data points align \textit{precisely} with the predicted scaling.

    \item \textbf{Over-Yield Rheological Master Curve} [Fig.~\ref{fig:S5}(b), Eq.~(\ref{eq:S_rate_collapse})]: \\
    By replacing $\Delta U_p$ with the single-cycle effective external work ratio ${W_{\text{ext}}}/{W_{\text{elastic}}} \propto {\dot{\gamma}^{0.17}}/{\gamma_A}$, both Region~3 and Region~4 collapse onto a universal master curve [Eq.~(\ref{eq:S_rate_collapse})]. This confirms that across over-yield regimes, the characteristic cycle count $N_c$ required to reach the peak potential energy is strictly governed by first-order rate-dependent rheological over-stress.

    \item \textbf{Fluctuation-Assisted Acceleration in Region 2} [Fig.~\ref{fig:S5}(b)]: \\
    In contrast, data points for Region~2 (sub-yield regime, $\gamma_A < \gamma_y$) fall systematically \textit{below} the master curve, leading to an overestimation of $N_c$ ($N_c^{\text{pred}} > N_c^{\text{MD}}$). In this sub-yield regime, the macroscopic first-order flow work ($W_{\text{ext}}$) is too weak to drive structural evolution alone. Instead, potential energy elevation is strongly accelerated by second-order localized elastic stress fluctuations ($\langle \delta\sigma_{\text{local}}^2 \rangle \propto \dot{\gamma}^{0.34}$)~\cite{Nicolas2018deformation}. These intense local fluctuations act as a ``mechanical catalyst,'' significantly lowering activation energy barriers for local shear transformation zones (STZs)~\cite{argon1979plastic} and enabling the system to reach its peak potential energy $\Delta U_p$ in far fewer cycles than predicted by macroscopic work scaling.
\end{enumerate}

\end{document}